\documentclass[sigconf]{acmart}

\AtBeginDocument{%
  }
\usepackage{url}
\usepackage{color}
\usepackage{lineno}
\usepackage{bm}
\usepackage{fancyvrb}
\usepackage{caption}
\usepackage{amsmath}
\usepackage{algorithmicx}
\usepackage{algorithm}
\usepackage{algpascal}
\usepackage{algc}
\usepackage{algcompatible}
\usepackage{algpseudocode}
\usepackage{listings}
\usepackage{cleveref}
\setcopyright{acmlicensed}
\copyrightyear{2018}
\acmYear{2018}
\acmDOI{XXXXXXX.XXXXXXX}

\acmConference[Conference acronym 'XX]{Make sure to enter the correct
  conference title from your rights confirmation email}{June 03--05,
  2018}{Woodstock, NY}
\acmISBN{978-1-4503-XXXX-X/2018/06}

\begin{document}

\title{DGEMM without FP64 Arithmetic -- Using FP64 Emulation and
FP8 Tensor Cores with Ozaki Scheme}

\author{Daichi Mukunoki}
\email{mukunoki@cc.nagoya-u.ac.jp}
\orcid{0000-0002-0051-6811}
\affiliation{%
  \institution{Information Technology Center, Nagoya University}
  \city{Nagoya}
  \state{Aichi}
  \country{Japan}
}
\renewcommand{\shortauthors}{D. Mukunoki}

\begin{abstract}
As the demand for AI computation rapidly increases, more hardware is being developed to efficiently perform the low-precision matrix multiplications required by such workloads. However, these operations are generally not directly applicable to scientific computations due to accuracy requirements. The Ozaki scheme -- an accurate matrix multiplication method proposed by Ozaki et al. in 2012 -- enables FP64 matrix multiplication (DGEMM) using low-precision matrix multiplication units, such as FP16 Tensor Cores. This approach has since been extended to utilize integer arithmetic, offering lower computational cost compared to floating-point-based implementations. In fact, it has achieved higher performance than hardware FP64 operations on GPUs equipped with fast INT8 Tensor Cores designed for AI workloads. However, recent AI-oriented processors trends have shifted toward improving the performance of low-precision floating-point operations, such as FP8, rather than integer operations. Motivated by this shift, this study revisits the use of low-precision floating-point operations in the Ozaki scheme. Specifically, we explore the use of FP8 Tensor Cores. In addition, for processors that support very slow or no hardware-based FP64 operations, we also consider FP64 arithmetic emulation based on integer arithmetic. This completely eliminates hardware FP64 instructions. Furthermore, we explore the use of blocking in the inner-product dimension to accelerate FP16-based implementations. We demonstrate the effectiveness of these methods by evaluating the performance on an NVIDIA RTX Blackwell architecture GPU. 
\end{abstract}

\begin{CCSXML}
<ccs2012>
 <concept>
  <concept_id>00000000.0000000.0000000</concept_id>
  <concept_desc>Do Not Use This Code, Generate the Correct Terms for Your Paper</concept_desc>
  <concept_significance>500</concept_significance>
 </concept>
 <concept>
  <concept_id>00000000.00000000.00000000</concept_id>
  <concept_desc>Do Not Use This Code, Generate the Correct Terms for Your Paper</concept_desc>
  <concept_significance>300</concept_significance>
 </concept>
 <concept>
  <concept_id>00000000.00000000.00000000</concept_id>
  <concept_desc>Do Not Use This Code, Generate the Correct Terms for Your Paper</concept_desc>
  <concept_significance>100</concept_significance>
 </concept>
 <concept>
  <concept_id>00000000.00000000.00000000</concept_id>
  <concept_desc>Do Not Use This Code, Generate the Correct Terms for Your Paper</concept_desc>
  <concept_significance>100</concept_significance>
 </concept>
</ccs2012>
\end{CCSXML}

%\ccsdesc[500]{Do Not Use This Code~Generate the Correct Terms for Your Paper}
%\ccsdesc[300]{Do Not Use This Code~Generate the Correct Terms for Your Paper}
%\ccsdesc{Do Not Use This Code~Generate the Correct Terms for Your Paper}
%\ccsdesc[100]{Do Not Use This Code~Generate the Correct Terms for Your Paper}

\keywords{Ozaki scheme, matrix multiplication, GEMM, Tensor Cores, FP8}


\maketitle

%========================================================
%========================================================
%========================================================
\section{Introduction}
\label{sec:introduction}
With the rapid advancement of AI technology, the demand for computing resources has surged. Since low-precision matrix multiplication is at the core of many AI workloads, processors with dedicated units like NVIDIA Tensor Cores have been developed. For example, on NVIDIA Blackwell architecture GPUs (2024), the 5th generation Tensor Core supports the following floating-point formats\footnote{FP32 (IEEE binary32) is supported only on the CUDA core.}: FP64 (IEEE binary64), TensorFloat-32 (TF32), FP16 (IEEE binary16), bfloat16 (BF16), FP8 (E4M3\footnote{ `E' and `M' indicate the bit lengths of the exponential and mantissa parts, respectively.} and E5M2), FP6 (E3M2 and E2M3), FP4 (E2M1), and the INT8 integer format. Furthermore, processors specifically designed for AI processing that do not support FP64 are also emerging.
\par %%%

However, traditional scientific computations that rely on FP64 cannot directly benefit from such hardware. The differing precision requirements between general-purpose and AI workloads increase hardware design complexity and raise development costs. Moreover, as AI workloads become more dominant in general-purpose systems, the improvement of FP64 performance is stagnating relative to that of low-precision operations. For instance, the performance ratio of FP64 to FP16, when using Tensor Cores and without sparsity, is 1:16 on the NVIDIA A100 (2020), approximately 1:29.5 on the H100 SXM (2022), and 1:125 on the GB200 NVL72 (2024). This stagnation challenges sustainable performance growth in scientific computing.
\par %%%

Matrix multiplication is a fundamental operation in scientific computing, widely used across various applications. It is computationally intensive, and its performance is highly dependent on FP64 arithmetic throughput. Consequently, matrix multiplication is among the operations most affected by the stagnation of FP64 performance improvements. In this paper, we refer to matrix multiplication as GEMM (general matrix multiplication), following the terminology of the Basic Linear Algebra Subprograms (BLAS) \cite{Dongarra:1990:SLB:77626.79170}.
\par %%%

The Ozaki scheme \cite{Ozaki:2012:ETM:2086820.2086827}, proposed by Ozaki et al. in 2012, is a method that enables high-precision matrix multiplication to be computed using low-precision matrix multiplications. For example, it allows FP64 GEMM (also known as DGEMM) to be performed using Tensor Core operations on FP16 data \cite{isc2020mukunoki}. Furthermore, this method has been extended to utilize integer GEMM \cite{ootomo2024dgemm}, which reduces computational cost compared to floating-point-based approaches. This technique has garnered significant attention due to its ability to exploit the rapidly improving performance of INT8 operations for AI workloads.
\par %%%

However, as of 2025, a noteworthy trend in AI processors is the shift from INT8 to lower-bit floating-point formats such as FP8. While INT8 is efficient for inference tasks \cite{jouppi2017tpu}, it is not well-suited for training. FP8, on the other hand, can be used for both inference and training \cite{micikevicius2022fp8formatsdeeplearning}, driving the development of unified FP8-based hardware. For example, as shown in Table~\ref{tab:gb_specs}\footnote{\url{https://resources.nvidia.com/en-us-blackwell-architecture}}, in the GB200 (2024) equipped with Blackwell architecture GPUs, FP8 and INT8 achieve the same throughput. However, in the planned specifications for the GB300, INT8 throughput is only about one-thirtieth that of FP8. Although the GB200 and GB300 may target different workloads, the emergence of a product that prioritizes FP8 over INT8 is noteworthy. For low precision INT, 
The Hopper architecture (2022) removed INT4 support from the Tensor Cores\footnote{CUDA cores, not Tensor Cores, support INT4\cite{10579250}.}, which had been introduced in the Ampere architecture (2020). As for AMD, the Instinct MI350 series (2025) supports INT8 with the same throughput as FP8. However, in the Keynote presentation at Advancing AI 2025\footnote{https://www.amd.com/content/dam/amd/en/documents/corporate/events/advancing-ai-2025-distribution-deck.pdf}, there was no mention of INT8 performance; instead, only the performance of low-precision FP was highlighted.
\par %%%

Against this background, we revisit the use of low-precision floating-point operations in the Ozaki scheme, targeting modern AI hardware. This paper first evaluates the performance of DGEMM using FP8 Tensor Cores on a Blackwell GPU, in comparison with FP16. Then, considering AI processors with limited or no FP64 support, we develop an implementation that eliminates FP64 instructions entirely and analyze its performance. Furthermore, we propose a new blocking strategy to further enhance performance. 
\par %%%

It should be noted that the primary objective of this study is not to outperform DGEMM implementations that utilize hardware FP64 units on specific processors. Rather, the goal is to offer new insights into the application of low-precision operations such as FP16 and FP8 in the Ozaki scheme for DGEMM emulation.
\par %%%

The remainder of this paper is organized as follows.
Section \ref{sec:related_work} introduces related work of this study.
Section \ref{sec:ozaki_scheme} outlines the overview of the Ozaki scheme.
Section \ref{sec:method} presents extended implementations of the Ozaki scheme.
Section \ref{sec:evaluation} presents the experimental results on an RTX Blackwell GPU.
Finally, the conclusion is presented in Section \ref{sec:conclusion}. 
\par %%%

\begin{table}[t]%%%%%%%%%%%%%%%%%%%%%%%%%%%%%%%%%%%%%%
\centering
\caption{Theoretical peak performance of GB300 NVL72 and GB200 NVL72 systems (Tensor Core performance does not consider sparsity).}
\label{tab:gb_specs}
\begin{tabular}{l|cc}
\hline
 & GB300 NVL72 & GB200 NVL72 \\\hline
INT8 Tensor Core & 12 POps/s & 360 POps/s \\
FP4 Tensor Core & 1,080 PFlops/s & 720 PFlops/s \\
FP8/FP6 Tensor Core & 360 PFlops/s & 360 PFlops/s \\
FP16/BF16 Tensor Core & 180 PFlops/s & 180 PFlops/s \\
TF32 Tensor Core & 90 PFlops/s & 90 PFlops/s \\
FP32 & 5760 TFlops/s & 5760 TFlops/s \\
FP64/FP64 Tensor Core & 100 TFlops/s & 2880 TFlops/s \\
\hline
\end{tabular}
\end{table}%%%%%%%%%%%%%%%%%%%%%%%%%%%%%%%%%%%%%%

%========================================================
%========================================================
%========================================================
\section{Related Work}
\label{sec:related_work}
The concept of performing high-precision computations using low-precision operations has existed for a long time. Early work includes double-word arithmetic proposed by Dekker in 1971 \cite{dekker1971}. It realizes double-length mantissa using floating-point arithmetic of a certain precision. This technique has been extended to triple-word \cite{10.1109/TC.2019.2918451} and quadruple-word \cite{930115} arithmetic. QD\footnote{https://github.com/BL-highprecision/QD} is known as an implementation of double- and quadruple-word arithmetic for Fortran and C++. These multi-word arithmetic techniques have been applied to enable FP64-equivalent computations on processors that either lack FP64 support or deliver significantly higher performance for lower-precision operations. For example, the use of double-word arithmetic to enhance precision on GPUs without native FP64 support has been explored \cite{10.1145/1179622.1179682}. Later, with the introduction of FP16 and BF16 \cite{8877390} for AI workloads, similar approaches were investigated. For instance, matrix multiplication using double-word arithmetic with FP16 \cite{doi:10.1137/21M1465032} and double- or triple-word arithmetic with BF16 \cite{8877427} have been studied. 
\par %%%

In addition to these floating-point arithmetic-based methods, high-precision (and arbitrary-precision) arithmetic libraries using integer arithmetic have been developed. For example, the GNU Multiple Precision Arithmetic Library (GMP)\footnote{https://gmplib.org} \cite{10.5555/2911024} is a C library that supports arbitrary-precision integer and floating-point arithmetic. The GNU MPFR Library (MPFR)\footnote{https://www.mpfr.org} \cite{mpfr} builds upon GMP and provides arbitrary-precision floating-point arithmetic compliant with the IEEE 754 standard, guaranteeing correct rounding. Also, GNU and Intel compilers provide FP128 (IEEE binary128, 128-bit floating-point, also known as quadruple-precision) arithmetic operations.
\par %%%

For GEMM, methods to achieve FP32-equivalent accuracy for GEMM using FP16 Tensor Cores \cite{8425458}\cite{doi:10.1177/10943420221090256} have been proposed. Later, the Ozaki scheme \cite{Ozaki:2012:ETM:2086820.2086827} has gained prominence as a method for computing high-precision GEMM using low-precision GEMM. It was originally proposed to compute accurate GEMM on matrices stored in a given floating-point format, using arithmetic of the same format. The method was later extended to allow different formats for data storage and computation. For example, it enables DGEMM to be computed using FP16 Tensor Cores \cite{isc2020mukunoki}. A major advantage of the Ozaki scheme is that it can utilize highly optimized GEMM routines provided for specific architectures, whereas previous methods require implementing GEMM from scratch. The scheme has also been applied in various contexts, such as high-precision GEMM (e.g., FP128 \cite{10.1145/3472456.3472493} and double-/triple-/quadruple-word arithmetic \cite{10.1007/978-3-031-37108-0_34}), computational methods ensuring reproducibility \cite{ozblas}\cite{10.1145/3432261.3432270}\cite{10.1007/978-3-031-30442-2_4}, and quantum chemistry calculations \cite{doi:10.1021/acs.jctc.4c00938}. 
\par %%%

A notable extension of the Ozaki scheme is its use of integer arithmetic proposed by Ootomo et al.\cite{ootomo2024dgemm} in 2024, which reduces computational cost compared to the floating-point-based approach. This is because the low-precision GEMMs in the Ozaki scheme are computed in fixed-point form and do not require handling the exponent part. This approach is well suited to the GPU architectures, as INT8 Tensor Core performance is rapidly improving. Subsequently, the integer-based method has been further refined \cite{abdelfattah2025analysisfloatingpointmatrixmultiplication}\cite{doi:10.1177/10943420241313064}. The latest advancement is Ozaki Scheme II \cite{ozaki2025ozakischemeiigemmoriented}, which leverages the Chinese Remainder Theorem (CRT) to further reduce computational cost.
\par %%%

Beyond GEMM, there are also efforts to apply low-precision operations to other computations. Examples include the use of FP16 Tensor Cores for Fast Fourier Transformations \cite{9555937} and the application of low-precision formats such as FP8 and FP16 in sparse matrix computations \cite{10.1007/978-3-031-69583-4_2}\cite{10793204}. Additionally, a comprehensive discussion on the effectiveness of AI-oriented matrix multiplication units, such as Tensor Cores, in various HPC applications has been presented \cite{9460517}.
\par %%%

%========================================================
%========================================================
\section{Ozaki Scheme}
\label{sec:ozaki_scheme}
This section outlines the mixed-precision version \cite{isc2020mukunoki} of the Ozaki scheme, which constitutes the core technique used in this study.
\par %%%

The Ozaki scheme is an error-free transformation for matrix multiplication that decomposes a single floating-point matrix multiplication into a sum of several matrix multiplications, each of which can be computed without rounding errors. Although this scheme was originally proposed for matrix multiplication, its fundamental concept can be regarded as an error-free transformation of an inner product. Therefore, we explain its principle here on the basis of inner products.
\par %%%

The scheme consists of the following three steps.
\begin{enumerate}
\renewcommand{\labelenumi}{Step.\arabic{enumi}. }
\setlength{\leftskip}{8pt}
\item \textbf{Slicing}\footnote{Some papers refer to this as ``splitting''.}: This step slices the input vectors into multiple vectors of the same length on an element-wise basis. This is done to ensure that the subsequent computation can be performed without rounding errors.
\item \textbf{Computation}: This step computes the products of all combinations of sliced vectors. At this stage, Tensor Cores can be utilized.
\item \textbf{Accumulation}\footnote{Some papers refer to this as ``summation''.}: This step accumulates the results obtained in the previous step to produce the final result.
\end{enumerate}
\par %%%

Hereafter, \texttt{TypeX} denotes a floating-point data type, such as FP64 or FP16. $\mathbb{F}_{\tt TypeX}$ represents the set of numbers representable in \texttt{TypeX}. ${\rm fl_{\tt TypeX}}(\cdot)$ and ${\rm cvt_{\tt TypeX}}(\cdot)$ denote the \texttt{TypeX} arithmetic operation and type-conversion to \texttt{TypeX} with round-to-nearest-even rounding, respectively. $u_{\tt TypeX}$ denotes the unit round-off of \texttt{TypeX}. $m_{\tt TypeX}$ is the number of mantissa bits of \texttt{TypeX}, including the hidden bit: $m_{\tt TypeX} = -\log_2 u_{\tt TypeX}$.
\par %%%

We now consider computing the inner product of $\bm{x}, \bm{y} \in {\mathbb{F}_{\tt Type1}}^{k}$ using a Tensor Core operation, which computes matrices stored in \texttt{Type2} with accumulation in \texttt{Type3}. First, we determine constants $\gamma$, $\xi$, and $\rho$ as defined below.

%%%%%%%%%%
% eq. (1)-(5)
%%%%%%%%%%
\begin{eqnarray}
\gamma &\coloneq& \left\lceil m_{\tt Type1} -  \frac{m_{\tt Type3} - \log_2 k)}{2}  \right\rceil \label{eq:gamma}\\
\xi &\coloneq& m_{\tt Type1} - m_{\tt Type2}\\
\rho &\coloneq& \max \left( \gamma, \xi \right) \label{eq:rho} 
\end{eqnarray}
\par %%%

Then, $\bm{x}$ in \texttt{Type1} is recursively sliced into $\underline{\bm{x}}^{(p)}_i$ in \texttt{Type2} as follows. Equations \eqref{eq:c}--\eqref{eq:xp} are processed recursively from $p=1$, starting with $\bm{x}^{(1)} = \bm{x}$, until $\bm{x}^{(p)} = 0$. The constant 0.75 in Equation \eqref{eq:sigma} was proposed by \cite{minamihata}. Equations \eqref{eq:xdash}--\eqref{eq:xp} are applied element-wise for $1 \leq i \leq k$. Here, we define that Equation \eqref{eq:c} yields ${c_x}^{(p)} \coloneq 0$ when $\max_{1 \leq i \leq k} {|{\bm{x}^{(p)}}_i|} = 0$.
%%%%%%%%%%
% eq. (3)-(6)
%%%%%%%%%%
\begin{eqnarray}
{c_x}^{(p)} &\coloneq& \left\lceil {\tt fl_{\tt Type1}} \left( \log_2 {\left( \max_{1\leq i \leq k}{\left|{\bm{x}^{(p)}_i}\right|}\right)} \right) \right\rceil \label{eq:c}\\
\sigma &\coloneq& 0.75\cdot 2^{\rho+{c_x}^{(p)}}  \label{eq:sigma}\\
{\bm{v}_i} &\coloneq& {\tt fl_{\tt Type1}} \left( \left({\bm{x}^{(p)}_i} + \sigma \right) - \sigma \right) \label{eq:xdash}\\
{\bm{x}^{(p+1)}_i} &\coloneq& {\tt fl_{\tt Type1}} \left( {\bm{x}^{(p)}_i} - {\bm{v}_i} \right) \\
{\underline{\bm{x}}^{(p)}_i} &\coloneq& {\rm cvt_{\tt Type2}}\left({\tt fl_{\tt Type1}} \left( 2^{-{c_x}^{(p)}} {\bm{v}_i} \right)\right) \label{eq:xp}
\end{eqnarray}

For Equation \eqref{eq:sigma}, if the condition $m_{\tt Type1} - \lfloor \log_{2} \sigma \rfloor > 0$ is not satisfied, the slicing process fails because $m_{\tt Type2}$ is insufficient to retain the information of a sliced vector.
\par %%%

Consequently, $\bm{x}$ is transformed as follows.
%%%%%%%%%%
% eq. (7)
%%%%%%%%%%
\begin{eqnarray}
\bm{x} = \sum_{p=1}^{s_x} 2^{{c_x}^{(p)}} \underline{\bm{x}}^{(p)}
\label{eq:split}
\end{eqnarray}
${c_x}^{(p)}$ corresponds to the exponent part of $\bm{x}$, and $\underline{\bm{x}}^{(p)}$ corresponds to the mantissa part of $\bm{x}$.
\par %%%

The vector $\bm{y}$ is processed in the same manner as $\bm{x}$. Consequently, the inner product $\bm{x}^T \bm{y}$ is transformed as follows:
%%%%%%%%%%
% eq. (8)
%%%%%%%%%%
\begin{eqnarray}
\bm{x}^{T}\bm{y} = \sum_{p=1}^{s_x} \sum_{q=1}^{s_y} 2^{{c_{x}}^{(p)}+{c_{y}}^{(q)}}{{\bm{\underline{x}}}^{(p)}}^T{\bm{\underline{y}}}^{(q)} \label{eq:oz}
\end{eqnarray}
In the above, the computation of the right-hand side can be performed using the \texttt{Type3} arithmetic without rounding errors. Therefore, the following equation is valid.
%%%%%%%%%%
% eq. (10)
%%%%%%%%%%
\begin{eqnarray}
{\underline{\bm{x}}^{(p)}}^T \underline{\bm{y}}^{(q)} = {\tt fl_{\tt Type3}} \left( {\underline{\bm{x}}^{(p)}}^T \underline{\bm{y}}^{(q)} \right)
\end{eqnarray}

It is important to note that in Equation \eqref{eq:oz}, the accumulation does not involve floating-point arithmetic; the equation holds exactly when performed without rounding errors. However, to achieve at least the accuracy of \texttt{Type1} arithmetic, accumulation can be performed using \texttt{Type1} arithmetic.
\par %%%

Although the above computation is described for inner product, it can be naturally extended to matrix multiplication. In this case, the computation in Equation \eqref{eq:oz} can be implemented using GEMM with Tensor Cores, as provided by cuBLAS\footnote{https://docs.nvidia.com/cuda/cublas/}.
\par %%%

The computational cost of this scheme depends on the number of slices, with memory consumption increasing accordingly. The number of slices is influenced by the following factors:
\begin{enumerate}
\renewcommand{\labelenumi}{\roman{enumi}. }
\item The inner dimension $k$ as in Equation \eqref{eq:gamma}.
\item The range of absolute values of the input elements, corresponding to the maximum in Equation \eqref{eq:c}.
\item The maximum number of significant bits in the input elements.  
\end{enumerate}
\par %%%

%========================================================
%========================================================
%========================================================
\section{Extended Implementation of Ozaki Scheme}
\label{sec:method}
This section proposes extended implementation methods for the existing Ozaki scheme in light of recent AI-oriented hardware developments. Two notable trends are observed: the introduction of low-precision floating-point formats such as FP8 replacing low-precision integer types like INT8, and the decline in FP64 performance, including the emergence of AI processors that lack FP64 support entirely. 
\par %%%

Our implementation builds upon OzBLAS\footnote{https://github.com/mukunoki/ozblas}, a DGEMM implementation using FP16 Tensor Cores developed in prior work \cite{isc2020mukunoki}. In addition, we incorporate several optimizations inspired by the FP128 GEMM implementation \cite{10.1145/3472456.3472493}. First, we improve accuracy by optimizing the order of accumulation. Second, since the Ozaki scheme performs matrix slicing along the inner-product direction, we transpose one of the input matrices in advance to enable continuous memory access, using \texttt{cublasDgeam} for the transposition. The sliced matrices are then processed by specifying the transposition mode in the GEMM routine.
\par %%%

%========================================================
%========================================================
\subsection{Using FP8 Tensor Cores}
\label{sec:tensor_cores}
This study targets NVIDIA RTX Blackwell architecture GPUs\footnote{https://images.nvidia.com/aem-dam/Solutions/geforce/blackwell/nvidia-rtx-blackwell-gpu-architecture.pdf} equipped with the 5th Generation Tensor Cores, specifically the GeForce RTX 5060 Ti. The theoretical floating-point performance is summarized in Table \ref{tab:gpus}.
\par %%%

This study focuses exclusively on FP8, since FP4 is too small to be practical in the Ozaki scheme and the target GPUs exhibit identical performance for FP6\footnote{As of cuBLAS 12.9, no GEMM routines supporting FP6 are available.} and FP8. However, for example, the AMD Instinct MI350X delivers FP6 performance that is twice that of FP8, suggesting that FP6 may be effective in certain environments. Among the FP8 formats, we adopt E4M3, which allows each slice to retain more mantissa bits than E5M2. The accumulation precision can be set to either FP32 or FP16, and we use FP32 accumulation, as higher accumulation precision tends to reduce the number of required slices. Computations are performed with \texttt{cublasGemmEx} in cuBLAS for FP16 and \texttt{cublasLtMatmul} in cuBLASLt for FP8. FP8 operations require 16-element memory alignment, and our implementation currently supports only problem sizes that are multiples of 16. This limitation could be addressed by introducing zero-padding, which we leave as future work.
\par %%%

Table \ref{tab:num_gemms} shows the minimum number of GEMMs required to compute DGEMMs using various Tensor Core operations, with respect to the inner dimension $k$. These values are for the case where all the FP64 mantissa bits are fully filled. The number of GEMMs corresponds to the square of the number of matrix slices. Columns marked with `--' indicate that slicing is no longer feasible due to insufficient mantissa precision, making the computation impossible. The case of FP8 (E5M2) is not shown here, but it leads to the same situation as FP6 (E3M2). In the Ozaki scheme, GEMM computation ideally dominates the overall execution time. Therefore, if a Tensor Core operation with FP8 input achieves twice the throughput of FP16 input while requiring twice as many GEMMs, the expected performance would be comparable between FP8 and FP16. However, it is important to note that when FP32 is used as the accumulation format, the difference in the number of GEMMs between FP16 and FP8 decreases as $k$ increases, giving FP8 the potential to outperform FP16.
\par %%%

\begin{table}[t]%%%%%%%%%%%%%%%%%%%%%%%%%%%%%%%%%%%%%%
\centering
\caption{Specification of GeForce RTX 5060 Ti GPU used in the evaluations. TC: Tensor Cores. Without sparsity. Performance is calculated based on the boost clock.}
\label{tab:gpus}
\begin{tabular}{l|r}
\hline
Specification & Value \\\hline
Clock (boost)	&		2.57 GHz	\\
Num of CUDA / Tensor Cores	&		4608 / 144	\\
Memory	&		GDDR7, 16 GB, 448 GB/s	\\
FP4TC (FP32 accum)	&		189.48 TFlops/s \\
FP6TC (FP32 accum)	&		94.74 TFlops/s \\
FP8TC (FP32 accum)	&		94.74 TFlops/s	\\
FP16TC (FP32 accum)	&		47.37 TFlops/s	\\
FP32 	&		23.69 TFlops/s	\\
FP64TC	&		0.3701 TFlops/s	\\
FP64	&		0.3701 TFlops/s	\\
\hline
\end{tabular}
\end{table}%%%%%%%%%%%%%%%%%%%%%%%%%%%%%%%%%%%%%%

\begin{table}[t]%%%%%%%%%%%%%%%%%%%%%%%%%%%%%%%%%%%%%%
\centering
\caption{Minimum number of GEMMs required for DGEMM using Tensor Cores. \texttt{Type2} is the input format and \texttt{Type3} is the accumulation format of Tensor Cores.}
\label{tab:num_gemms}
\begin{tabular}{r|rrrrrr}
\hline
Type2&	FP16	&	FP16	&	FP8	&	FP8	&	FP6	&	FP6	\\
&		&		&	(E4M3)	&	(E4M3)	&	(E3M2)	&	(E3M2)	\\\hline
Type3&	FP32	&	FP16	&	FP32	&	FP16	&	FP32	&	FP16	\\\hline
$k=$ 8	&	25	&	121	&	121	&	121	&	196	&	196	\\
16	&	25	&	196	&	121	&	196	&	196	&	196	\\
32	&	36	&	196	&	121	&	196	&	196	&	196	\\
64	&	36	&	324	&	121	&	324	&	196	&	324	\\
128	&	36	&	324	&	121	&	324	&	196	&	324	\\
256	&	36	&	729	&	121	&	729	&	196	&	729	\\
512	&	49	&	729	&	121	&	729	&	196	&	729	\\
1024	&	49	&	2809	&	121	&	2809	&	196	&	2809	\\
2048	&	64	&	2809	&	121	&	2809	&	196	&	2809	\\
4096	&	64	&	--	&	121	&	--	&	196	&	--	\\
8192	&	81	&	--	&	121	&	--	&	196	&	--	\\
16384	&	81	&	--	&	121	&	--	&	196	&	--	\\
32768	&	121	&	--	&	121	&	--	&	196	&	--	\\
65536	&	121	&	--	&	121	&	--	&	196	&	--	\\
131072	&	196	&	--	&	196	&	--	&	196	&	--	\\
262144	&	196	&	--	&	196	&	--	&	196	&	--	\\
\hline
\end{tabular}
\end{table}%%%%%%%%%%%%%%%%%%%%%%%%%%%%%%%%%%%%%%

%========================================================
%========================================================
\subsection{FP64 Arithmetic Emulation}
\label{sec:fp64_elimination}
When computing DGEMM using the Ozaki scheme, some FP64 operations are required for the slicing and accumulation processes. Specifically, it requires addition, multiplication, max, and a comparison operator ($a < b$). 
\par%%%

We develop an implementation that excludes hardware FP64 instructions, assuming processors with no hardware support for FP64 operations. Our approach emulates FP64 operations using 32-bit and 64-bit integer arithmetic, in a manner similar to MPFR. Since integer operations are typically required for address calculations, we assume that performance degradation is unlikely even on AI-oriented processors. 
\par%%%

The emulation algorithms for addition and multiplication are conceptually related to Dekker’s double-length floating-point algorithm. Specifically, the mantissa of an FP64 value is split into two 32-bit integer values, and the upper and lower parts are computed in a manner analogous to pen-and-paper arithmetic with two-digit numbers. The results are then computed using 64-bit integer arithmetic. Listing~\ref{lst:fp64emuMull} shows part of the FP64 multiplication code. The exponent is computed separately. Our implementation employs round-to-nearest-even rounding, producing results that are bitwise identical to hardware FP64 operations. However, for simplicity, it omits handling overflow, underflow, and subnormal numbers.
\par%%%

An alternative approach is to use a three-word extension of Dekker's algorithm with FP32 arithmetic. This approach has been used to speed up FP128 GEMM with FP64 using the Ozaki scheme \cite{10.1145/3472456.3472493}. However, this technique still requires some FP64 operations, and the exponent range is limited to that of FP32. Therefore, this study explores the integer-based approach.
\par%%%

\begin{lstlisting}[caption = Part of the FP64 multiplication code, label = lst:fp64emuMull]
typedef struct {
    uint32_t parts[4];
} uint32x4;

__device__ inline uint32x4 mul_mantissa (uint64_t a, uint64_t b) {
    uint32_t a_low = a & 0xFFFFFFFF;
    uint32_t a_high = a >> 32;
    uint32_t b_low = b & 0xFFFFFFFF;
    uint32_t b_high = b >> 32;

    uint64_t p00 = (uint64_t)a_low * b_low;
    uint64_t p01 = (uint64_t)a_low * b_high;
    uint64_t p10 = (uint64_t)a_high * b_low;
    uint64_t p11 = (uint64_t)a_high * b_high;

    uint64_t middle = p01 + p10;
    uint64_t carry = (middle < p01) ? (1ULL << 32) : 0;

    uint64_t low_result = p00 + ((middle & 0xFFFFFFFF) << 32);
    carry += (low_result < p00) ? 1 : 0;
    uint64_t high_result = p11 + (middle >> 32) + carry;

    uint32x4 result;
    result.parts[0] = low_result & 0xFFFFFFFF;
    result.parts[1] = low_result >> 32;
    result.parts[2] = high_result & 0xFFFFFFFF;
    result.parts[3] = high_result >> 32;
    return result;
}
\end{lstlisting}

%========================================================
%========================================================
\subsection{Inner-product-wise blocking}
\label{sec:k-blocking}
As shown in Table \ref{tab:num_gemms}, the smaller the precision gap between \texttt{Type2} and \texttt{Type3}, the more rapidly the number of GEMMs increases with respect to the inner dimension $k$. To address this issue, the matrix can be partitioned along the inner product dimension, the Ozaki scheme can be applied to each block, and the results can then be accumulated. In this paper, we refer to this strategy as inner-product-wise blocking. Since this approach applies error-free transformation on a block-by-block basis, numerical errors are introduced during the accumulation of the block results, leading to lower accuracy compared to the non-blocking case. However, if the accumulation is performed using FP64 arithmetic, the resulting accuracy is comparable to that of a GEMM computed in FP64 arithmetic. Therefore, this loss of accuracy is not problematic when the goal is to emulate DGEMM.
\par %%%

In addition, this blocking approach contributes to a reduction in the working memory required to store slice matrices, which is one of the drawbacks in the Ozaki scheme. Prior studies have addressed this issue by applying blocking along the outer dimension, a strategy we refer to in this paper as outer-product-wise blocking. In practice, it is feasible to combine both inner-product-wise and outer-product-wise blocking to execute computations within constrained memory budgets.
\par %%%

Furthermore, inner-product-wise blocking has a minor advantage that it avoids redundant slicing of the same region in one matrix with outer-product-wise blocking. However, this overhead is practically negligible, as slice processing accounts for only a small fraction of the total execution time, as demonstrated in Figure~\ref{fig:break} in Section~\ref{sec:evaluation}. Nonetheless, this issue should still be taken into account depending on the shape and size of the matrices.
\par %%%

However, the primary concern with inner-product-wise blocking is that each block computation updates the entire result matrix, leading to increased memory accesses. Furthermore, a common drawback of both blocking strategies is the potential degradation in execution efficiency (throughput) as the block size decreases.
\par %%%

Considering the above, the trade-off between advantages and disadvantages must be considered in determining the block sizes, which are tuning parameters. In this study, we do not implement an automatic mechanism for determining optimal block sizes; instead, we manually set a suitable block size for the sole purpose of demonstrating the concept of inner-product-wise blocking.
\par %%%

%========================================================
%========================================================
%========================================================
\section{Evaluation}
\label{sec:evaluation}
This section demonstrates the performance of the implementation described in the previous section.
\par %%%

For the performance of DGEMM with the Ozaki scheme, the following abbreviations are used hereafter.
\begin{itemize}
    \item \textbf{DGEMM-FP8TC}: DGEMM using FP8 Tensor Cores with FP32 accumulation.
    \item \textbf{DGEMM-FP16TC}: DGEMM using FP16 Tensor Cores with FP32 accumulation.
    \item \textbf{FP64emu}: using the FP64 arithmetic emulation in slicing and accumulation.
\end{itemize}

\subsection{Experimental Setting}
We conducted evaluations on Geforce RTX 5060 Ti (RTX Blackwell architecture) with 16 GB of memory, with CUDA 12.9 (nvcc V12.9.86). Table \ref{tab:gpus} summarizes the specification of the GPU. We used the best value from three executions. Work memory is allocated all at once before the computation, and that time is not included in performance. 
\par %%%

We compute $C = AB$, where matrix A is $m \times k$ and matrix B is $k \times n$. Those matrices are square matrices ($m = n = k$). We initialized the input FP64 matrices with pseudo-uniform random numbers of (1, 10) to have the same number of GEMMs shown in Table \ref{tab:num_gemms}. Considering that GEMM is $O(n^3)$ and the other processes are $O(n^2)$, this experimental condition is considered to be the case where the cost of GEMMs as a percentage of overall execution time is minimized.
\par %%%

\subsection{Performance of cuBLAS GEMMs}
\begin{figure}[t]
\centering
\includegraphics[width=\hsize]{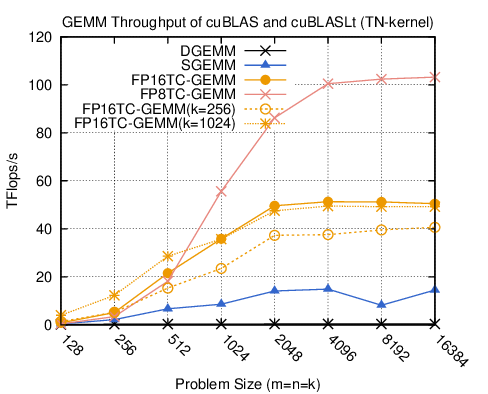}
\caption{GEMM throughput of cuBLAS and cuBLASLt using various formats. Matrix A is transposed (TN-kernel). If $k$ is specified, it is fixed across all problem sizes.}
\label{fig:gemm}
\end{figure}

First, as a preliminary evaluation, we measured the performance of the GEMM routines used in the DGEMM emulation of the Ozaki scheme, along with the standard DGEMM and SGEMM routines for reference. Figure~\ref{fig:gemm} shows the throughput of GEMM for each format. The evaluated GEMM routines are as follows.
\begin{itemize}
    \item \textbf{DGEMM}: \texttt{cublasDgemm}
    \item \textbf{SGEMM}: \texttt{cublasSgemm}
    \item \textbf{FP16TC-GEMM}: \texttt{cublasGemmEx} using FP16 Tensor Cores with FP32 accumulation
    \item \textbf{FP8TC-GEMM}: \texttt{cublasLtMatmul} using FP8 Tensor Cores with FP32 accumulation
\end{itemize}
To optimize memory access in the slicing process, matrix $A$ is transposed, as mentioned in Section~\ref{sec:method}; thus, this performance evaluation corresponds to the case where matrix $A$ is transposed (the so-called TN kernel). When the problem size is sufficiently large, the performance approaches the theoretical peak performance\footnote{Depending on runtime conditions, the GPU may operate at frequencies exceeding the boost clock, resulting in observed performance surpassing the theoretical peak.}. The results with fixed $k$ are intended to evaluate throughput when the block size is set to $k$ in inner-product-wise blocking. For FP16, as shown in Table \ref{tab:num_gemms}, the number of GEMMs is 36 when $k = 256$, but increases to 49 for $k = 512$ and $1024$, and to 64 for $k = 2048$. There is almost no drop in throughput for $k = 1024$, but a decrease is observed for $k = 256$. Therefore, from the perspective of GEMM throughput and the number of GEMMs, adopting a block size of 1024 is optimal. For FP8, the number of GEMMs remains constant within this problem size range.
\par %%%

%==================================
%==================================

%==================================
\begin{figure}[t]
\centering
\includegraphics[width=\hsize]{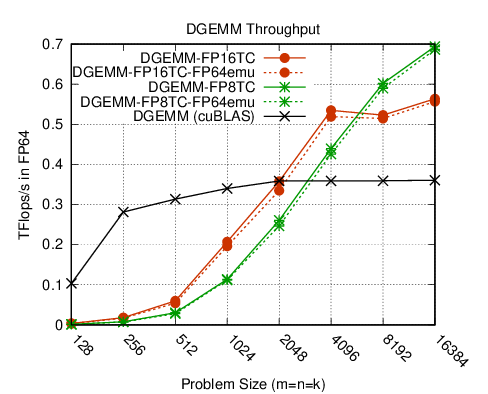}
\caption{DGEMM throughput in FP64-equivalent Flops/s. Inner-product-wise
blocking is not applied. Work-memory size is 8 GB.}
\label{fig:perf}
\end{figure}
%==================================

\subsection{Accuracy of DGEMM using Ozaki Scheme}
\label{sec:accuracy}
Our implementation in this paper achieves an error level equivalent to or lower than that of standard triple-loop GEMM with FP64 arithmetic, regardless of whether FP8 or FP16 Tensor Cores are used or not. This is not specific to our implementation, but a general nature of the Ozaki scheme when using all slices of matrices and using the GEMM results with the square of the number of slices (e.g., as shown in \cite{10.1145/3472456.3472493}). In the Ozaki scheme, no rounding error occurs during the GEMM computation phase. Rounding error arises only during the accumulation phase with FP64 arithmetic. 
\par %%%

\begin{figure}[t]
\centering
\includegraphics[width=\hsize]{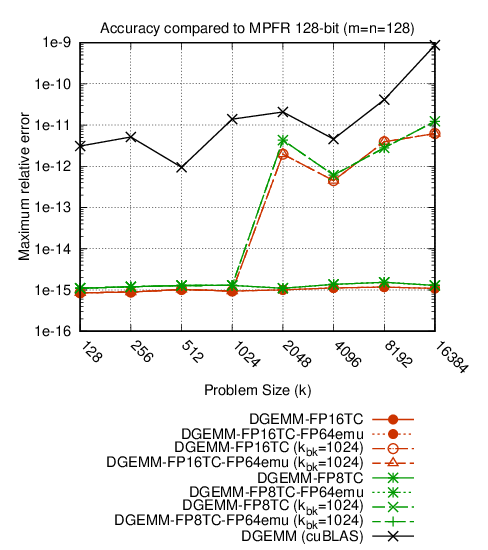}
\caption{Accuracy comparison with MPFR 128-bit arithmetic. If $k_{bk}$ is specified, the inner-product-wise blocking with $k_{bk}$ is used.}
\label{fig:accuracy}
\end{figure}

Figure \ref{fig:accuracy} shows the maximum relative error compared to the GEMM computed with 128-bit arithmetic by MPFR. The MPFR result is rounded to FP64. Using FP8 increases the number of slices compared to FP16, resulting in more additions being performed. Consequently, the accumulation of rounding errors may be slightly larger. Furthermore, utilizing inner-product-wise blocking increases the number of accumulations, further increasing a larger accumulation of rounding errors. The FP64 arithmetic emulation is performed using round-to-nearest-even rounding, producing results that are bitwise identical to those of hardware FP64 operations.
\par %%%

%==================================
%==================================

\subsection{Performance of DGEMM using Ozaki Scheme}
\begin{figure}[t]
\centering

\begin{minipage}{0.45\hsize}
\includegraphics[width=\hsize]{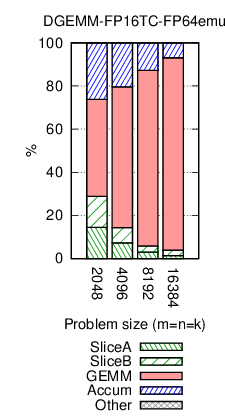}
\end{minipage}
\begin{minipage}{0.45\hsize}
\includegraphics[width=\hsize]{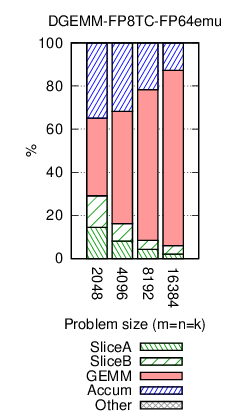}
\end{minipage}

\caption{Execution time breakdown. Inner-product-wise blocking is not applied. Work-memory size is 8 GB.}
\label{fig:break}
\end{figure}

Figure~\ref{fig:perf} shows the performance results. Inner-product-wise blocking is not applied in this experiment. Note that the Flops/s values are expressed in FP64-equivalent units, calculated based on the number of floating-point operations performed by DGEMM in FP64 arithmetic ($2m^3$), rather than the actual number of operations executed by Tensor Cores. Figure~\ref{fig:break} presents the performance breakdown.
\par %%%

DGEMM-FP8TC outperforms DGEMM-FP16TC for $m \geq 8192$. This is because the throughput of FP8 Tensor Cores is twice that of FP16 Tensor Cores, whereas the number of GEMMs at $m=8192$ is only approximately $121/81 \approx 1.49$ times greater for DGEMM-FP8TC than for DGEMM-FP16TC, as shown in Table~\ref{tab:num_gemms}. At $m=16384$, DGEMM-FP8TC is approximately 1.26 times faster, which is consistent with the expected speedup, $(94.74/121) / (47.37/81) \approx 1.34$, based solely on the number of GEMMs. Although DGEMM-FP8TC is theoretically expected to be faster even at $m=2048$ -- where the number of GEMMs is about 1.89 times greater -- the breakdown reveals that non-GEMM overhead becomes significant at this point, and GEMM count alone no longer determines overall performance. 
\par %%%

When the problem size is small, the performance of DGEMM emulation using the Ozaki scheme falls significantly below that of cuBLAS DGEMM. As shown in Figure~\ref{fig:gemm}, this occurs because FP8TC-GEMM and FP16TC-GEMM require larger problem sizes to reach their theoretical peak performance. The underlying reason is that the Bytes/Flop ratio decreases as precision decreases. This issue may be alleviated to some extent by using batched GEMM \cite{ozblas}.
\par %%%

For the FP64 arithmetic emulation, the performance degradation is modest. The slicing and accumulation processes are memory-intensive operations. Moreover, when the problem size is sufficiently large, the time spent on slicing and accumulation constitutes only a small fraction of the total execution time. Consequently, the cost of FP64 arithmetic operations does not become apparent.
\par %%%

\begin{figure}[t]
\centering

\begin{minipage}{0.45\hsize}
\includegraphics[width=\hsize]{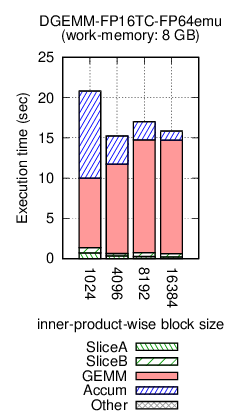}
\end{minipage}
\begin{minipage}{0.45\hsize}
\includegraphics[width=\hsize]{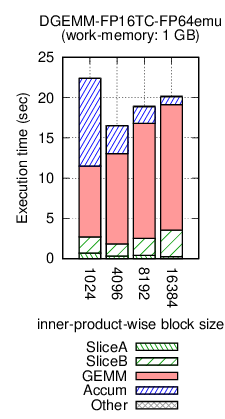}
\end{minipage}

\caption{Execution time when $m=n=k=16384$ with various inner-product-wise block size. Work-memory size is 8 GB (left) or 1 GB (right).}
\label{fig:kbk}
\end{figure}

Figure~\ref{fig:kbk} shows the execution time and its breakdown for various inner-product-wise block sizes in DGEMM-FP16TC when the working memory is 8 GB or 1 GB. For the block size $k_{bk}$, the number of GEMMs is 49 for $k_{bk}=1024$, 64 for $k_{bk}=4096$, and 81 for $k_{bk}=16384$, as listed in Table~\ref{tab:num_gemms}. While the GEMM cost clearly decreases with decreasing $k_{bk}$, the accumulation cost increases. Consequently, the optimal block size is $k_{bk}=4096$ in both cases. 
\par %%%

%========================================================
%========================================================
%========================================================
\section{Conclusion}
\label{sec:conclusion}
In this paper, we examined extensions to an existing Ozaki scheme implementation to compute DGEMM using low-precision floating-point GEMMs, motivated by recent trends in AI-oriented processors. We evaluated the following points on the RTX 5060 Ti GPU based on the RTX Blackwell architecture.
\par%%%

First, we considered the use of FP8 Tensor Cores, which are becoming standard in AI computation and replacing INT8. We demonstrated that the use of FP8 Tensor Cores is feasible. Their performance is competitive with that of FP16 Tensor Cores when FP8 achieves roughly twice the throughput of FP16. Moreover, for processors that support very slow FP64 operations or do not support them at all, we considered the use of the FP64 arithmetic emulation based on integer arithmetic. Only modest performance degradation was observed, because the slicing and accumulation processes are memory-intensive. We also explored the inner-product-wise blocking for performance improvement when using Tensor Core operations where the input and computation precision are close. Specifically, this was introduced with the expectation that it would mitigate the weakness of FP16 Tensor Cores with FP32 accumulation compared to FP8 Tensor Cores with FP32 accumulation. Although its effectiveness was limited due to the significant impact of increased accumulation costs, performance improvements were observed under certain conditions.
\par%%%

A variety of improvements for the Ozaki scheme have been proposed, and performance evaluation using combinations of these methods remains future work. Several ideas have various advantages-disadvantages tradeoffs, and comprehensive performance evaluation is required. 
\par%%%

The growing demand for AI computing has led to the introduction of AI-oriented designs in general-purpose processors as well as AI-specialized processors. However, the requirements for computational precision differ between AI and traditional scientific computations, making the integration of these hardware architectures increasingly challenging. Moreover, as demand for AI computing continues to expand, there is concern that the role of FP64 will decline further relative to the growing reliance on low-precision operations. Achieving FP64-equivalent precision using low-precision operations through the Ozaki scheme or classical high-precision arithmetic methods is expected to mitigate these difficulties to some extent. At the same time, the computational precision and hardware requirements for AI computing are changing rapidly, necessitating the development of emulation technologies that address these trends. To sustain performance improvements in scientific computing, a multifaceted approach encompassing AI technology trends, hardware design, and FP64 emulation techniques will be required.
\par%%%

%\textcolor{red}{The source code used in writing this paper is available here\footnote{https://github.com/XXXXXX} (to be added after acceptance).}

%==================================
%==================================

%==================================
%==================================

\begin{acks}
%\textcolor{red}{Acknowledgments will be added here after acceptance.} 
This work was supported by JSPS KAKENHI Grant Number JP25K24387. 
\end{acks}

\bibliographystyle{ACM-Reference-Format}
\bibliography{arxiv_hpcasia2026}


\begin{thebibliography}{32}

%%% the Web address.

\ifx \showCODEN    \undefined \def \showCODEN     #1{\unskip}     \fi
\ifx \showISBNx    \undefined \def \showISBNx     #1{\unskip}     \fi
\ifx \showISBNxiii \undefined \def \showISBNxiii  #1{\unskip}     \fi
\ifx \showISSN     \undefined \def \showISSN      #1{\unskip}     \fi
\ifx \showLCCN     \undefined \def \showLCCN      #1{\unskip}     \fi
\ifx \shownote     \undefined \def \shownote      #1{#1}          \fi
\ifx \showarticletitle \undefined \def \showarticletitle #1{#1}   \fi
\ifx \showURL      \undefined \def \showURL       {\relax}        \fi
% invisible to TeX
\providecommand\bibfield[2]{#2}
\providecommand\bibinfo[2]{#2}
\providecommand\natexlab[1]{#1}
\providecommand\showeprint[2][]{arXiv:#2}

\bibitem[Abdelfattah et~al\mbox{.}(2025)]%
        {abdelfattah2025analysisfloatingpointmatrixmultiplication}
\bibfield{author}{\bibinfo{person}{Ahmad Abdelfattah}, \bibinfo{person}{Jack Dongarra}, \bibinfo{person}{Massimiliano Fasi}, \bibinfo{person}{Mantas Mikaitis}, {and} \bibinfo{person}{Françoise Tisseur}.} \bibinfo{year}{2025}\natexlab{}.
\newblock \bibinfo{title}{Analysis of Floating-Point Matrix Multiplication Computed via Integer Arithmetic}.
\newblock
\showeprint[arxiv]{2506.11277}~[math.NA]
\urldef\tempurl%
\url{https://arxiv.org/abs/2506.11277}
\showURL{%
\tempurl}


\bibitem[Burgess et~al\mbox{.}(2019)]%
        {8877390}
\bibfield{author}{\bibinfo{person}{Neil Burgess}, \bibinfo{person}{Jelena Milanovic}, \bibinfo{person}{Nigel Stephens}, \bibinfo{person}{Konstantinos Monachopoulos}, {and} \bibinfo{person}{David Mansell}.} \bibinfo{year}{2019}\natexlab{}.
\newblock \showarticletitle{Bfloat16 Processing for Neural Networks}. In \bibinfo{booktitle}{\emph{2019 IEEE 26th Symposium on Computer Arithmetic (ARITH)}}. \bibinfo{pages}{88--91}.
\newblock
\href{https://doi.org/10.1109/ARITH.2019.00022}{doi:\nolinkurl{10.1109/ARITH.2019.00022}}


\bibitem[Dawson et~al\mbox{.}(2024)]%
        {doi:10.1021/acs.jctc.4c00938}
\bibfield{author}{\bibinfo{person}{William Dawson}, \bibinfo{person}{Katsuhisa Ozaki}, \bibinfo{person}{Jens Domke}, {and} \bibinfo{person}{Takahito Nakajima}.} \bibinfo{year}{2024}\natexlab{}.
\newblock \showarticletitle{Reducing Numerical Precision Requirements in Quantum Chemistry Calculations}.
\newblock \bibinfo{journal}{\emph{Journal of Chemical Theory and Computation}} \bibinfo{volume}{20}, \bibinfo{number}{24} (\bibinfo{year}{2024}), \bibinfo{pages}{10826--10837}.
\newblock
\href{https://doi.org/10.1021/acs.jctc.4c00938}{doi:\nolinkurl{10.1021/acs.jctc.4c00938}}
\newblock
\shownote{PMID: 39644230}.


\bibitem[Dekker(1971)]%
        {dekker1971}
\bibfield{author}{\bibinfo{person}{T.~J. Dekker}.} \bibinfo{year}{1971}\natexlab{}.
\newblock \showarticletitle{{{A Floating-Point Technique for Extending the Available Precision}}}.
\newblock \bibinfo{journal}{\emph{Numer. Math.}}  \bibinfo{volume}{18} (\bibinfo{year}{1971}), \bibinfo{pages}{224--242}.
\newblock


\bibitem[Domke et~al\mbox{.}(2021)]%
        {9460517}
\bibfield{author}{\bibinfo{person}{Jens Domke}, \bibinfo{person}{Emil Vatai}, \bibinfo{person}{Aleksandr Drozd}, \bibinfo{person}{Peng ChenT}, \bibinfo{person}{Yosuke Oyama}, \bibinfo{person}{Lingqi Zhang}, \bibinfo{person}{Shweta Salaria}, \bibinfo{person}{Daichi Mukunoki}, \bibinfo{person}{Artur Podobas}, \bibinfo{person}{Mohamed WahibT}, {and} \bibinfo{person}{Satoshi Matsuoka}.} \bibinfo{year}{2021}\natexlab{}.
\newblock \showarticletitle{Matrix Engines for High Performance Computing: A Paragon of Performance or Grasping at Straws?}. In \bibinfo{booktitle}{\emph{2021 IEEE International Parallel and Distributed Processing Symposium (IPDPS)}}. \bibinfo{pages}{1056--1065}.
\newblock
\href{https://doi.org/10.1109/IPDPS49936.2021.00114}{doi:\nolinkurl{10.1109/IPDPS49936.2021.00114}}


\bibitem[Dongarra et~al\mbox{.}(1990)]%
        {Dongarra:1990:SLB:77626.79170}
\bibfield{author}{\bibinfo{person}{J.~J. Dongarra}, \bibinfo{person}{C.~J. Du}, \bibinfo{person}{S. Hammarling}, {and} \bibinfo{person}{I.~S. Duff}.} \bibinfo{year}{1990}\natexlab{}.
\newblock \showarticletitle{{A Set of Level 3 Basic Linear Algebra Subprograms}}.
\newblock \bibinfo{journal}{\emph{ACM Trans. Math. Softw.}} \bibinfo{volume}{16}, \bibinfo{number}{1} (\bibinfo{date}{March} \bibinfo{year}{1990}), \bibinfo{pages}{1--17}.
\newblock


\bibitem[Fabiano et~al\mbox{.}(2019)]%
        {10.1109/TC.2019.2918451}
\bibfield{author}{\bibinfo{person}{Nicolas Fabiano}, \bibinfo{person}{Jean-Michel Muller}, {and} \bibinfo{person}{Joris Picot}.} \bibinfo{year}{2019}\natexlab{}.
\newblock \showarticletitle{Algorithms for Triple-Word Arithmetic}.
\newblock \bibinfo{journal}{\emph{IEEE Trans. Comput.}} \bibinfo{volume}{68}, \bibinfo{number}{11} (\bibinfo{date}{Nov.} \bibinfo{year}{2019}), \bibinfo{pages}{1573–1583}.
\newblock
\showISSN{0018-9340}
\href{https://doi.org/10.1109/TC.2019.2918451}{doi:\nolinkurl{10.1109/TC.2019.2918451}}


\bibitem[Fasi et~al\mbox{.}(2023)]%
        {doi:10.1137/21M1465032}
\bibfield{author}{\bibinfo{person}{Massimiliano Fasi}, \bibinfo{person}{Nicholas~J. Higham}, \bibinfo{person}{Florent Lopez}, \bibinfo{person}{Theo Mary}, {and} \bibinfo{person}{Mantas Mikaitis}.} \bibinfo{year}{2023}\natexlab{}.
\newblock \showarticletitle{Matrix Multiplication in Multiword Arithmetic: Error Analysis and Application to GPU Tensor Cores}.
\newblock \bibinfo{journal}{\emph{SIAM Journal on Scientific Computing}} \bibinfo{volume}{45}, \bibinfo{number}{1} (\bibinfo{year}{2023}), \bibinfo{pages}{C1--C19}.
\newblock
\href{https://doi.org/10.1137/21M1465032}{doi:\nolinkurl{10.1137/21M1465032}}


\bibitem[Fousse et~al\mbox{.}(2007)]%
        {mpfr}
\bibfield{author}{\bibinfo{person}{L. Fousse}, \bibinfo{person}{G. Hanrot}, \bibinfo{person}{V. Lef{\`e}vre}, \bibinfo{person}{P. P{\'e}lissier}, {and} \bibinfo{person}{P. Zimmermann}.} \bibinfo{year}{2007}\natexlab{}.
\newblock \showarticletitle{{MPFR: A Multiple-precision Binary Floating-point Library with Correct Rounding}}.
\newblock \bibinfo{journal}{\emph{ACM Trans. Math. Software}} \bibinfo{volume}{33}, \bibinfo{number}{2} (\bibinfo{year}{2007}), \bibinfo{pages}{13:1--13:15}.
\newblock


\bibitem[Graillat et~al\mbox{.}(2024)]%
        {10.1007/978-3-031-69583-4_2}
\bibfield{author}{\bibinfo{person}{Stef Graillat}, \bibinfo{person}{Fabienne J{\'e}z{\'e}quel}, \bibinfo{person}{Theo Mary}, \bibinfo{person}{Rom{\'e}o Molina}, {and} \bibinfo{person}{Daichi Mukunoki}.} \bibinfo{year}{2024}\natexlab{}.
\newblock \showarticletitle{Reduced-Precision and Reduced-Exponent Formats for Accelerating Adaptive Precision Sparse Matrix--Vector Product}. In \bibinfo{booktitle}{\emph{Euro-Par 2024: Parallel Processing}}, \bibfield{editor}{\bibinfo{person}{Jesus Carretero}, \bibinfo{person}{Sameer Shende}, \bibinfo{person}{Javier Garcia-Blas}, \bibinfo{person}{Ivona Brandic}, \bibinfo{person}{Katzalin Olcoz}, {and} \bibinfo{person}{Martin Schreiber}} (Eds.). \bibinfo{publisher}{Springer Nature Switzerland}, \bibinfo{address}{Cham}, \bibinfo{pages}{17--30}.
\newblock


\bibitem[Granlund and Team(2015)]%
        {10.5555/2911024}
\bibfield{author}{\bibinfo{person}{Torbjrn Granlund} {and} \bibinfo{person}{Gmp~Development Team}.} \bibinfo{year}{2015}\natexlab{}.
\newblock \bibinfo{booktitle}{\emph{GNU MP 6.0 Multiple Precision Arithmetic Library}}.
\newblock \bibinfo{publisher}{Samurai Media Limited}, \bibinfo{address}{London, GBR}.
\newblock
\showISBNx{9789888381968}


\bibitem[Henry et~al\mbox{.}(2019)]%
        {8877427}
\bibfield{author}{\bibinfo{person}{Greg Henry}, \bibinfo{person}{Ping Tak~Peter Tang}, {and} \bibinfo{person}{Alexander Heinecke}.} \bibinfo{year}{2019}\natexlab{}.
\newblock \showarticletitle{Leveraging the bfloat16 Artificial Intelligence Datatype For Higher-Precision Computations}. In \bibinfo{booktitle}{\emph{2019 IEEE 26th Symposium on Computer Arithmetic (ARITH)}}. \bibinfo{pages}{69--76}.
\newblock
\href{https://doi.org/10.1109/ARITH.2019.00019}{doi:\nolinkurl{10.1109/ARITH.2019.00019}}


\bibitem[Hida et~al\mbox{.}(2001)]%
        {930115}
\bibfield{author}{\bibinfo{person}{Y. Hida}, \bibinfo{person}{X.S. Li}, {and} \bibinfo{person}{D.H. Bailey}.} \bibinfo{year}{2001}\natexlab{}.
\newblock \showarticletitle{Algorithms for quad-double precision floating point arithmetic}. In \bibinfo{booktitle}{\emph{Proceedings 15th IEEE Symposium on Computer Arithmetic. ARITH-15 2001}}. \bibinfo{pages}{155--162}.
\newblock
\href{https://doi.org/10.1109/ARITH.2001.930115}{doi:\nolinkurl{10.1109/ARITH.2001.930115}}


\bibitem[Jouppi et~al\mbox{.}(2017)]%
        {jouppi2017tpu}
\bibfield{author}{\bibinfo{person}{Norman~P Jouppi}, \bibinfo{person}{Cliff Young}, \bibinfo{person}{Nishant Patil}, \bibinfo{person}{David Patterson}, \bibinfo{person}{Gaurav Agrawal}, \bibinfo{person}{Raminder Bajwa}, \bibinfo{person}{Sarah Bates}, \bibinfo{person}{Suresh Bhatia}, \bibinfo{person}{Nan Boden}, \bibinfo{person}{Al Borchers}, {et~al\mbox{.}}} \bibinfo{year}{2017}\natexlab{}.
\newblock \showarticletitle{In-datacenter performance analysis of a tensor processing unit}. In \bibinfo{booktitle}{\emph{Proceedings of the 44th Annual International Symposium on Computer Architecture}}. ACM, \bibinfo{pages}{1--12}.
\newblock
\href{https://doi.org/10.1145/3079856.3080246}{doi:\nolinkurl{10.1145/3079856.3080246}}


\bibitem[Kouya and Utsugiri(2023)]%
        {10.1007/978-3-031-37108-0_34}
\bibfield{author}{\bibinfo{person}{Tomonori Kouya} {and} \bibinfo{person}{Taiga Utsugiri}.} \bibinfo{year}{2023}\natexlab{}.
\newblock \showarticletitle{Optimization of Multiple-Precision LU Decomposition Using Ozaki Scheme}. In \bibinfo{booktitle}{\emph{Computational Science and Its Applications -- ICCSA 2023 Workshops}}, \bibfield{editor}{\bibinfo{person}{Osvaldo Gervasi}, \bibinfo{person}{Beniamino Murgante}, \bibinfo{person}{Ana Maria A.~C. Rocha}, \bibinfo{person}{Chiara Garau}, \bibinfo{person}{Francesco Scorza}, \bibinfo{person}{Yeliz Karaca}, {and} \bibinfo{person}{Carmelo~M. Torre}} (Eds.). \bibinfo{publisher}{Springer Nature Switzerland}, \bibinfo{address}{Cham}, \bibinfo{pages}{529--545}.
\newblock
\showISBNx{978-3-031-37108-0}


\bibitem[Li et~al\mbox{.}(2021)]%
        {9555937}
\bibfield{author}{\bibinfo{person}{Binrui Li}, \bibinfo{person}{Shenggan Cheng}, {and} \bibinfo{person}{James Lin}.} \bibinfo{year}{2021}\natexlab{}.
\newblock \showarticletitle{tcFFT: A Fast Half-Precision FFT Library for NVIDIA Tensor Cores}. In \bibinfo{booktitle}{\emph{2021 IEEE International Conference on Cluster Computing (CLUSTER)}}. \bibinfo{pages}{1--11}.
\newblock
\href{https://doi.org/10.1109/Cluster48925.2021.00035}{doi:\nolinkurl{10.1109/Cluster48925.2021.00035}}


\bibitem[Luo et~al\mbox{.}(2024)]%
        {10579250}
\bibfield{author}{\bibinfo{person}{Weile Luo}, \bibinfo{person}{Ruibo Fan}, \bibinfo{person}{Zeyu Li}, \bibinfo{person}{Dayou Du}, \bibinfo{person}{Qiang Wang}, {and} \bibinfo{person}{Xiaowen Chu}.} \bibinfo{year}{2024}\natexlab{}.
\newblock \showarticletitle{{ Benchmarking and Dissecting the Nvidia Hopper GPU Architecture }}. In \bibinfo{booktitle}{\emph{2024 IEEE International Parallel and Distributed Processing Symposium (IPDPS)}}. \bibinfo{publisher}{IEEE Computer Society}, \bibinfo{address}{Los Alamitos, CA, USA}, \bibinfo{pages}{656--667}.
\newblock
\href{https://doi.org/10.1109/IPDPS57955.2024.00064}{doi:\nolinkurl{10.1109/IPDPS57955.2024.00064}}


\bibitem[Markidis et~al\mbox{.}(2018)]%
        {8425458}
\bibfield{author}{\bibinfo{person}{Stefano Markidis}, \bibinfo{person}{Steven Wei~Der Chien}, \bibinfo{person}{Erwin Laure}, \bibinfo{person}{Ivy~Bo Peng}, {and} \bibinfo{person}{Jeffrey~S. Vetter}.} \bibinfo{year}{2018}\natexlab{}.
\newblock \showarticletitle{NVIDIA Tensor Core Programmability, Performance \& Precision}. In \bibinfo{booktitle}{\emph{2018 IEEE International Parallel and Distributed Processing Symposium Workshops (IPDPSW)}}. \bibinfo{pages}{522--531}.
\newblock
\href{https://doi.org/10.1109/IPDPSW.2018.00091}{doi:\nolinkurl{10.1109/IPDPSW.2018.00091}}


\bibitem[Micikevicius et~al\mbox{.}(2022)]%
        {micikevicius2022fp8formatsdeeplearning}
\bibfield{author}{\bibinfo{person}{Paulius Micikevicius}, \bibinfo{person}{Dusan Stosic}, \bibinfo{person}{Neil Burgess}, \bibinfo{person}{Marius Cornea}, \bibinfo{person}{Pradeep Dubey}, \bibinfo{person}{Richard Grisenthwaite}, \bibinfo{person}{Sangwon Ha}, \bibinfo{person}{Alexander Heinecke}, \bibinfo{person}{Patrick Judd}, \bibinfo{person}{John Kamalu}, \bibinfo{person}{Naveen Mellempudi}, \bibinfo{person}{Stuart Oberman}, \bibinfo{person}{Mohammad Shoeybi}, \bibinfo{person}{Michael Siu}, {and} \bibinfo{person}{Hao Wu}.} \bibinfo{year}{2022}\natexlab{}.
\newblock \bibinfo{title}{FP8 Formats for Deep Learning}.
\newblock
\showeprint[arxiv]{2209.05433}~[cs.LG]
\urldef\tempurl%
\url{https://arxiv.org/abs/2209.05433}
\showURL{%
\tempurl}


\bibitem[Minamihata et~al\mbox{.}(2016)]%
        {minamihata}
\bibfield{author}{\bibinfo{person}{A. Minamihata}, \bibinfo{person}{K. Ozaki}, \bibinfo{person}{T. Ogita}, {and} \bibinfo{person}{S. Oishi}.} \bibinfo{year}{2016}\natexlab{}.
\newblock \showarticletitle{Improved extraction scheme for accurate floating-point summation}. In \bibinfo{booktitle}{\emph{The 35th JSST Annual Conference International Conference on Simulation Technology}}.
\newblock


\bibitem[Mukunoki et~al\mbox{.}(2020a)]%
        {ozblas}
\bibfield{author}{\bibinfo{person}{Daichi Mukunoki}, \bibinfo{person}{Takeshi Ogita}, {and} \bibinfo{person}{Katsuhisa Ozaki}.} \bibinfo{year}{2020}\natexlab{a}.
\newblock \showarticletitle{{Reproducible BLAS Routines with Tunable Accuracy Using Ozaki Scheme for Many-core Architectures}}. In \bibinfo{booktitle}{\emph{Proc.~13th International Conference on Parallel Processing and Applied Mathematics (PPAM2019), Lecture Notes in Computer Science}}, Vol.~\bibinfo{volume}{12043}. \bibinfo{publisher}{Springer Berlin Heidelberg}, \bibinfo{pages}{516--527}.
\newblock
\href{https://doi.org/10.1007/978-3-030-43229-4_44}{doi:\nolinkurl{10.1007/978-3-030-43229-4_44}}


\bibitem[Mukunoki et~al\mbox{.}(2021a)]%
        {10.1145/3432261.3432270}
\bibfield{author}{\bibinfo{person}{Daichi Mukunoki}, \bibinfo{person}{Katsuhisa Ozaki}, \bibinfo{person}{Takeshi Ogita}, {and} \bibinfo{person}{Roman Iakymchuk}.} \bibinfo{year}{2021}\natexlab{a}.
\newblock \showarticletitle{Conjugate Gradient Solvers with High Accuracy and Bit-Wise Reproducibility between CPU and GPU Using Ozaki Scheme}. In \bibinfo{booktitle}{\emph{The International Conference on High Performance Computing in Asia-Pacific Region}} (Virtual Event, Republic of Korea) \emph{(\bibinfo{series}{HPC Asia 2021})}. \bibinfo{publisher}{Association for Computing Machinery}, \bibinfo{address}{New York, NY, USA}, \bibinfo{pages}{100–109}.
\newblock
\showISBNx{9781450388429}
\href{https://doi.org/10.1145/3432261.3432270}{doi:\nolinkurl{10.1145/3432261.3432270}}


\bibitem[Mukunoki et~al\mbox{.}(2020b)]%
        {isc2020mukunoki}
\bibfield{author}{\bibinfo{person}{Daichi Mukunoki}, \bibinfo{person}{Katsuhisa Ozaki}, \bibinfo{person}{Takeshi Ogita}, {and} \bibinfo{person}{Toshiyuki Imamura}.} \bibinfo{year}{2020}\natexlab{b}.
\newblock \showarticletitle{{DGEMM using Tensor Cores, and Its Accurate and Reproducible Versions}}. In \bibinfo{booktitle}{\emph{ISC High Performance 2020, Lecture Notes in Computer Science}}, Vol.~\bibinfo{volume}{12151}. \bibinfo{publisher}{Springer International Publishing}, \bibinfo{pages}{230--248}.
\newblock
\href{https://doi.org/10.1007/978-3-030-50743-5_12}{doi:\nolinkurl{10.1007/978-3-030-50743-5_12}}


\bibitem[Mukunoki et~al\mbox{.}(2021b)]%
        {10.1145/3472456.3472493}
\bibfield{author}{\bibinfo{person}{Daichi Mukunoki}, \bibinfo{person}{Katsuhisa Ozaki}, \bibinfo{person}{Takeshi Ogita}, {and} \bibinfo{person}{Toshiyuki Imamura}.} \bibinfo{year}{2021}\natexlab{b}.
\newblock \showarticletitle{Accurate Matrix Multiplication on Binary128 Format Accelerated by Ozaki Scheme}. In \bibinfo{booktitle}{\emph{Proceedings of the 50th International Conference on Parallel Processing}} (Lemont, IL, USA) \emph{(\bibinfo{series}{ICPP '21})}. \bibinfo{publisher}{Association for Computing Machinery}, \bibinfo{address}{New York, NY, USA}, Article \bibinfo{articleno}{78}, \bibinfo{numpages}{11}~pages.
\newblock
\showISBNx{9781450390682}
\href{https://doi.org/10.1145/3472456.3472493}{doi:\nolinkurl{10.1145/3472456.3472493}}


\bibitem[Mukunoki et~al\mbox{.}(2023)]%
        {10.1007/978-3-031-30442-2_4}
\bibfield{author}{\bibinfo{person}{Daichi Mukunoki}, \bibinfo{person}{Katsuhisa Ozaki}, \bibinfo{person}{Takeshi Ogita}, {and} \bibinfo{person}{Toshiyuki Imamura}.} \bibinfo{year}{2023}\natexlab{}.
\newblock \showarticletitle{Infinite-Precision Inner Product and Sparse Matrix-Vector Multiplication Using Ozaki Scheme with Dot2 on Manycore Processors}. In \bibinfo{booktitle}{\emph{Parallel Processing and Applied Mathematics}}, \bibfield{editor}{\bibinfo{person}{Roman Wyrzykowski}, \bibinfo{person}{Jack Dongarra}, \bibinfo{person}{Ewa Deelman}, {and} \bibinfo{person}{Konrad Karczewski}} (Eds.). \bibinfo{publisher}{Springer International Publishing}, \bibinfo{address}{Cham}, \bibinfo{pages}{40--54}.
\newblock
\showISBNx{978-3-031-30442-2}


\bibitem[Ootomo et~al\mbox{.}(2024)]%
        {ootomo2024dgemm}
\bibfield{author}{\bibinfo{person}{Hiroyuki Ootomo}, \bibinfo{person}{Katsuhisa Ozaki}, {and} \bibinfo{person}{Rio Yokota}.} \bibinfo{year}{2024}\natexlab{}.
\newblock \showarticletitle{DGEMM on integer matrix multiplication unit}.
\newblock \bibinfo{journal}{\emph{The International Journal of High Performance Computing Applications}} (\bibinfo{year}{2024}).
\newblock
\href{https://doi.org/10.1177/10943420241239588}{doi:\nolinkurl{10.1177/10943420241239588}}


\bibitem[Ootomo and Yokota(2022)]%
        {doi:10.1177/10943420221090256}
\bibfield{author}{\bibinfo{person}{Hiroyuki Ootomo} {and} \bibinfo{person}{Rio Yokota}.} \bibinfo{year}{2022}\natexlab{}.
\newblock \showarticletitle{{Recovering single precision accuracy from Tensor Cores while surpassing the FP32 theoretical peak performance}}.
\newblock \bibinfo{journal}{\emph{The International Journal of High Performance Computing Applications}} \bibinfo{volume}{36}, \bibinfo{number}{4} (\bibinfo{year}{2022}), \bibinfo{pages}{475--491}.
\newblock
\href{https://doi.org/10.1177/10943420221090256}{doi:\nolinkurl{10.1177/10943420221090256}}


\bibitem[Ozaki et~al\mbox{.}(2012)]%
        {Ozaki:2012:ETM:2086820.2086827}
\bibfield{author}{\bibinfo{person}{K. Ozaki}, \bibinfo{person}{T. Ogita}, \bibinfo{person}{S. Oishi}, {and} \bibinfo{person}{S.~M. Rump}.} \bibinfo{year}{2012}\natexlab{}.
\newblock \showarticletitle{{Error-free transformations of matrix multiplication by using fast routines of matrix multiplication and its applications}}.
\newblock \bibinfo{journal}{\emph{Numer. Algorithms}} \bibinfo{volume}{59}, \bibinfo{number}{1} (\bibinfo{year}{2012}), \bibinfo{pages}{95--118}.
\newblock


\bibitem[Ozaki et~al\mbox{.}(2025)]%
        {ozaki2025ozakischemeiigemmoriented}
\bibfield{author}{\bibinfo{person}{Katsuhisa Ozaki}, \bibinfo{person}{Yuki Uchino}, {and} \bibinfo{person}{Toshiyuki Imamura}.} \bibinfo{year}{2025}\natexlab{}.
\newblock \bibinfo{title}{Ozaki Scheme II: A GEMM-oriented emulation of floating-point matrix multiplication using an integer modular technique}.
\newblock
\showeprint[arxiv]{2504.08009}~[cs.MS]
\urldef\tempurl%
\url{https://arxiv.org/abs/2504.08009}
\showURL{%
\tempurl}


\bibitem[Thall(2006)]%
        {10.1145/1179622.1179682}
\bibfield{author}{\bibinfo{person}{Andrew Thall}.} \bibinfo{year}{2006}\natexlab{}.
\newblock \showarticletitle{Extended-precision floating-point numbers for GPU computation}. In \bibinfo{booktitle}{\emph{ACM SIGGRAPH 2006 Research Posters}} (Boston, Massachusetts) \emph{(\bibinfo{series}{SIGGRAPH '06})}. \bibinfo{publisher}{Association for Computing Machinery}, \bibinfo{address}{New York, NY, USA}, \bibinfo{pages}{52–es}.
\newblock
\showISBNx{1595933646}
\href{https://doi.org/10.1145/1179622.1179682}{doi:\nolinkurl{10.1145/1179622.1179682}}


\bibitem[Uchino et~al\mbox{.}(2025)]%
        {doi:10.1177/10943420241313064}
\bibfield{author}{\bibinfo{person}{Yuki Uchino}, \bibinfo{person}{Katsuhisa Ozaki}, {and} \bibinfo{person}{Toshiyuki Imamura}.} \bibinfo{year}{2025}\natexlab{}.
\newblock \showarticletitle{Performance enhancement of the Ozaki Scheme on integer matrix multiplication unit}.
\newblock \bibinfo{journal}{\emph{The International Journal of High Performance Computing Applications}} \bibinfo{volume}{39}, \bibinfo{number}{3} (\bibinfo{year}{2025}), \bibinfo{pages}{462--476}.
\newblock
\href{https://doi.org/10.1177/10943420241313064}{doi:\nolinkurl{10.1177/10943420241313064}}


\bibitem[Yang et~al\mbox{.}(2024)]%
        {10793204}
\bibfield{author}{\bibinfo{person}{Dechuang Yang}, \bibinfo{person}{Yuxuan Zhao}, \bibinfo{person}{Yiduo Niu}, \bibinfo{person}{Weile Jia}, \bibinfo{person}{En Shao}, \bibinfo{person}{Weifeng Liu}, \bibinfo{person}{Guangming Tan}, {and} \bibinfo{person}{Zhou Jin}.} \bibinfo{year}{2024}\natexlab{}.
\newblock \showarticletitle{Mille-feuille: A Tile-Grained Mixed Precision Single-Kernel Conjugate Gradient Solver on GPUs}. In \bibinfo{booktitle}{\emph{SC24: International Conference for High Performance Computing, Networking, Storage and Analysis}}. \bibinfo{pages}{1--16}.
\newblock
\href{https://doi.org/10.1109/SC41406.2024.00064}{doi:\nolinkurl{10.1109/SC41406.2024.00064}}


\end{thebibliography}

\end{document}